\newcommand{\simle}{\mbox{$\stackrel{<}{_{\sim}}$}}
\newcommand{\simge}{\mbox{$\stackrel{>}{_{\sim}}$}}

\documentstyle[aaspp4,flushrt]{article}
\lefthead{Monnier et al.}
\righthead{Spectra of late-type stars}

\begin{document}
%_____________________________________TITLE PAGE___________________________
\title{Temporal Variations of Mid-IR Spectra in Late-Type Stars}
\author{J. D. Monnier\altaffilmark{1}, T. R. Geballe\altaffilmark{2}, 
and W. C. Danchi\altaffilmark{1}
} 
\altaffiltext{1}{Space Sciences Laboratory, University of California, Berkeley,
Berkeley,  CA  94720-7450
}
%\author{T. R. Geballe}
\altaffiltext{2}{Joint Astronomy Centre,
600 North A'ohoku Place, University Park, Hilo, HI 96720}
%
%
%
%
%____________________________________ABSTRACT PAGE_________________________
\begin{abstract}
New multi-epoch, mid-infrared (8-13\,$\micron$) spectrophotometric
observations are presented for 30~late-type stars.  The observations
were collected over a four year period (1994-1997), permitting an
investigation of the mid-infrared spectral shape as a function of the
pulsation cycle (typically 1-2 years).  The spectra of stars with
little excess infrared emission and those with carbon-rich dust show
the least spectral variability, while stars with evidence for dusty,
oxygen-rich envelopes are most likely to show discernible variations in
their spectral profile. Most significantly, a large fraction of
variable stars with strong 9.7\,$\micron$ emission features show clear
spectral profile changes which repeat from one cycle to the next.  The
significant sharpening of the silicate feature near maximum light can
not be fully explained by heating and cooling of the circumstellar dust
shell during the pulsational cycle, suggesting that the dust optical
properties themselves must also be varying.  In addition, the
appearance of a narrow emission feature near the silicate peak for a
few stars may require the production of especially ``pure'' silicate
dust near maximum light.  The general narrowing of the silicate feature
observed may reflect the evolution of the pre-existing
dirty grains whose surface impurities have been evaporated off when the
grain temperature rises preceding maximum light.  An improved theory of
dust formation which can explain the observed changes in the grain
properties around a single, pulsating star may lead to a definitive
explanation for the diversity of silicate emission profiles observed
amongst oxygen-rich, late-type stars.

\end{abstract}

\keywords{stars: AGB and post-AGB, stars: circumstellar matter, 
stars: mass-loss, stars: variables, ISM: dust }

%_______________________________________INTRODUCTION_______________________
\section{Introduction}
The mid-infrared (8-13\,$\micron$) spectra of late-type stars have been
measured by many observers since the development of infrared
detectors.  These red giants and supergiants are often surrounded by
dusty envelopes which absorb stellar radiation and re-radiate the
energy in the near- and mid-infrared.  The infrared spectra can be
classified based on the chemical content of the circumstellar
environment (oxygen- or carbon-rich) and on the optical thickness of
the dusty envelope (e.g., Merrill \& Stein 1976a,b).  Oxygen-rich
circumstellar environments often produce spectra evincing a feature
near 9.7\,$\micron$ resulting from the presence of silicate dust (Woolf
\& Ney 1969).  This feature appears in emission for optically thin
envelopes or in absorption when large enough optical depths are
encountered.  The emission spectra of dust surrounding carbon stars are
nearly featureless, although often containing an 11.3\,$\micron$
feature attributed to SiC.  In addition, many of these red giants and
supergiants are classified as long-period variables, pulsating with a
typical period of 1-2 years.

The homogeneous set of survey measurements by the Infra-Red
Astronomical Satellite (IRAS) in the mid-1980s allowed observers to
classify silicate emission features based on various schemes (IRAS
Science Team 1986; Little-Marenin \& Little 1988, 1990; Goebel et al.
1989; Sloan \& Price 1995).  The different shapes of the feature have
been interpreted largely as due to differences in the chemical make-up
of oxygen-rich dust.  Unfortunately the IRAS program did not
conscientiously include observations of the mid-infrared spectra of
long-period variable stars at different phases of their luminosity
cycles, and there are only a few cases where such data have been
retrieved from the IRAS Low Resolution Spectrometer (LRS) database.
These observations have suggested silicate feature strength variations
as a function of pulsational cycle, but have been hampered by limited
temporal coverage (Little-Marenin, Stencel, \& Staley 1996).  More
recent results by Creech-Eakman et al. (1997) point towards evidence
for variations in the silicate feature as a function of pulsational
phase, but the comparison spectra were taken nearly a decade apart.
Hence, the simple observational question of whether the mid-infrared
spectra of LPVs change shape through the pulsational cycle has been
left without a decisive answer.

A campaign of observations taken from 1994 to 1997 was designed to
monitor the mid-infrared spectrum of nearly 30 late-type stars.  The
observations,  sampling the spectrum of most stars multiple times
within a pulsational cycle, used the same instrument and observing
technique. The homogeneity of this data set is important for allowing
reliable spectral comparisons, avoiding complicating issues such as
different apertures and calibration methods.  This paper presents the
full data set collected thus far and discusses the spectral variability
(or lack of variability) of our sample stars.

%____________________________PRESENT_OBSERVATIONS_______________________
\section{Observations}

Mid-infrared spectrophotometry was carried out with the United Kingdom
Infra-Red Telescope (UKIRT) from 1994 to 1997.  These observations
employed the 32-element, linear array spectrometer CGS3 and were
obtained with a 5$\arcsec$ diameter aperture and standard chopping and
nodding techniques.  Wavelength calibrations were derived from
observations of a krypton arc lamp in 4th, 5th, and 6th order, and are
accurate to $\pm$0.02\,$\micron$.  The wavelength resolution of the
observations was $\sim$0.2\,$\micron$, sampled three times per
resolution element.  Flux calibrations were derived from observations
of $\alpha$~Lyr, $\alpha$~Aur, $\alpha$~CMa, and other bright standard
stars.  All standard stars used for ratioing are of spectral type K0 or
earlier, so that absorption in the fundamental band of SiO, which is
very prominent in late K and M giants and supergiants, is minimal in
the ratioed spectra and does not affect the shapes of the reduced
spectra.

The uncertainty in the {\em absolute} flux level was determined from
the internal consistency of an intercomparison of photometric standards
observed on a given night. Specific determinations of the uncertainty
for individual nights, along with other observing information, can be
found in Table~1.  Note that for spectra taken on 1994 November~20,
which was not photometric, the uncertainty is probably $\pm$20\%.

Intracomparison of spectra from stars with little dust emission taken
on different nights ($\alpha$~Boo, $\delta$~Oph, etc.) indicates the
{\em relative} flux calibration within a single spectrum to be accurate
to about one percent.  Details of the spectral shape are least
reliable near 9.7\,$\micron$, due to strong telluric absorption by
ozone.  Inspection of the entire data set reveals the relative
calibration for 1997 March~17 to be less accurate than the others,
occasionally showing a 5-15\% miscalibration longward of 12\,$\micron$
presumably due to changing telluric H$_2$O absorption.  It is important
to refer to Table~1 when judging the quality of an individual
spectrum.

%_____________________________________RESULTS__________________________
\section{Results}

The observed stars were divided into three categories after inspecting
the full set of spectra.  Figure~1 shows spectra from 18~stars whose
mid-infrared spectral shapes showed no apparent changes during our
observing campaign; this includes stars which were observed
only a single time ($\beta$~Peg, Egg~Nebula, and U~Her).  Figure~2
contains spectra from 8 stars whose 9.7\,$\micron$ silicate feature
show clear enhancement over the stellar/dust continuum during maximum
light.  Lastly, there are 4~stars which belong to neither of the above
groups and these spectra can be found in Figure~3.  These stars show
either a change in the mid-infrared spectral slope or increased rms
fluctuations in the spectral shape.

Figures~1-3 all share the same format.  The upper panels show the full
set of calibrated data, while the lower panels display the relative
spectra for each source as determined by the following method.  First,
normalized spectra are produced by dividing each spectrum by its mean
flux between 8-13\,$\micron$ and then are smoothed ($\Delta \lambda =
0.2\,\micron$).  Then, after removing spectra with known calibration
problems (1997 March~17 observations and the cloudy data from 1994
August~26, see Table 1), the set of normalized spectra for each star is
averaged together to form a mean spectral shape.  Finally, this mean
spectrum is divided into each of the normalized spectra to produce
plots of the deviation from the cycle-averaged spectral shape, which
are displayed in the lower panels.  The rms fluctuation about the mean
spectral shape is calculated for each star and is denoted by $\sigma$
each lower panel.  Spectra deemed unreliable are not included here and
thus the bottom panel is absent if our observations lacked a 
sufficient number of ``clean'' spectra (at least two).

In addition, each panel contains a legend which tabulates the mean
8-13\,$\micron$ fluxes in Janskys, the dates of observation, and the
pulsation cycle phases, where applicable.  Table~2 contains the
pulsational characteristics, mostly drawn from the recently released
Hipparcos Catalog (ESA 1997), along with spectral type and the
number of spectra presented in this work.  A few stars which are
classified as semiregulars have their pulsational phases included only
if a very recent phase determination is available (e.g., VX~Sgr and
W~Hya).  Each date of observation is assigned its own linestyle for all
figures to facilitate intercomparisons between different stars and
different nights.  The vertical scale of the bottom panels is fixed so
the relative magnitude of the spectral shape variations within
this sample can be easily accessed.  Although the data presented here do not
have adequate temporal coverage to justify a detailed analysis of the
light curves, inspection confirms the earlier finding that the infrared
maximum is generally about $\Delta \phi \approx 0.1$ after the optical
maximum for Mira variables (Lockwood \& Wing 1971).  

\subsection{``Constant'' Stars}

Figure~1 contains the observations of stars whose spectra showed no
discernible shape changes during this campaign.  Some of these stars
are both spectrally constant and show little excess infrared radiation,
justifying their traditional role as calibrators for spectral
observations.  In particular, $\alpha$~Boo, $\alpha$~Her, $\alpha$~Tau,
and $\delta$~Oph are among the most spectrally constant of all objects
observed, and were used to infer the true calibration uncertainty in
this survey.   Only stars having spectral shape variations less than
2.0\% rms and no overall slope changes were selected to be included as
``constant'' stars.  Note again that the 1994 August~26 absolute photometry
for $\alpha$~Boo, $\alpha$~Sco, and $\delta$~Oph is not reliable due to
clouds (see Table 1).

Of all the strongly variable stars with clear 9.7\,$\micron$ silicate
features included in this survey, only NML~Cyg appears to possess a
constant spectral shape in this wavelength regime.  Despite a
significant change in flux during this campaign
($\sim$0.25~magnitudes), the spectral shape is remarkably constant.  By
considering multi-wavelength photometric and interferometric
measurements, Monnier et al. (1997) estimated the 11\,$\micron$ optical
depth to be about 2; hence, it is not surprising that the spectral
shape hardly changes, since the dust at the $\tau_{\rm mid-IR}=1$
surface lies far enough away from the star as to not to be strongly
affected by its varying luminosity.  These same arguments apply for the
relatively flat mid-IR spectrum of VY~CMa, whose optical depth at
11\,$\micron$ has also been estimated to be $\simge 2$ (Danchi et al.
1994; Le Sidaner \& Le Bertre 1996).  Be reminded, however, that
interpretations of present interferometric data is subject to change,
since most of it has implicitly assumed spherical symmetry.  The
situation is further complicated for VY~CMa, because it's mid-IR
spectrum has apparently recently undergone a dramatic change.  About 25
years ago, Merrill \& Stein (1976a,b) observed this silicate feature to
be in strong emission (similar in strength to VX~Sgr), and we can not
offer an explanation for its ``disappearance.''

The University of California at Berkeley Infrared Spatial
Interferometer (ISI) survey of the 11\,$\micron$ angular sizes of late-type
stars (Danchi et al. 1994) detected two stars whose dust shells were
quite distant from the stellar surface ($R_{\rm dust}\simge 40
R_\star$), $\alpha$~Ori and $\alpha$~Sco.  While no firm conclusions can
be drawn for $\alpha$~Sco due to a lack of ``clean'' spectra, 
sufficient high-quality $\alpha$~Ori spectra were available 
to conclude that its mid-infrared spectral shape was constant to within
error bars.  This constancy is consistent with the interferometric
observations, which suggest the dust is located far from the star,
relatively immune to the effects of stellar variability.  However,
a recent report (Bester et al. 1996) indicates a fresh dust formation
episode around $\alpha$~Ori, contributing $\sim$10\% of the
11\,$\micron$ flux.  Modest changes in this hot dust spectrum 
during our campaign of observations would not be definitively
detected within our experimental uncertainty since the flux from the
cool, distant dust shell dominates the mid-IR dust emission.

By inspecting the deviations from the cycle-averaged spectral shapes
(the bottom panels of Figure~1), one can see many bumps and dips
resulting from miscalibration.  No firm conclusion in this paper has
been based on any fluctuations which are both small (few percent) and
not correlated with the pulsational phase (or flux) of the star.  These
fluctuations limit the precision to which the peak of the spectral
features can be measured and should be not interpreted beyond the
uncertainty implied by these observations of standard stars.

In summary, stars with weak silicate features or no features at all are
most likely to show no spectral shape variations.  In addition, not
one carbon stars in our sample (CIT~6, IRC~+10216, or V~Hya) showed
an overall change of spectral shape or variation in the strength of 
the SiC feature during a luminosity cycle (refer to
\S3.3 for IRC~+10216 spectra).  For some O-rich stars, mid-infrared
interferometric observations exist to correlate little spectral
variability with either high dust shell opacity or the presence of a dust 
shell far from the central star ($R_{\rm dust} \simge 40 R_\star$).

\subsection{Silicate Feature Variation}

The most interesting results of this paper are found in Figure~2, in
which data are presented for 8 stars which exhibit systematic
spectral changes directly related to the pulsational phase.  It is
worth noting that a large fraction of the surveyed stars with a
distinct 9.7\,$\micron$ feature are in this category.  Two different
types of spectral shape variability are observed: an overall narrowing
of the silicate feature near maximum light seen in most of the stars
and a more narrow emission feature near the peak of the silicate
feature seen only for a few sources.  Even CIT~3, whose spectrum
reveals a dusty envelope with optical depth only slightly smaller than
that for NML~Cyg (as judged by the partial self-absorption of the
silicate feature), shows the same systematic shape changes as for stars
with optically thin envelopes (e.g., $o$~Cet and U~Ori).

The certain identification of the spectral changes as related to the
pulsational phase is confirmed in two different ways.  First, the
reality of this effect is attested to by the large number of stars
which show the sharpening of the 9.7\,$\micron$ silicate feature near
maximum.  There is no case where the silicate peak is seen to grow
relative to the dust emission in the wings (near 12\,$\micron$, for
example) at the minimum of the luminosity cycle.  A second way to
positively correlate the spectral changes with the flux output of the
star is to obtain spectra at the same phase or flux level, but one
pulsational period apart.  In cases where this proved possible
($\chi$~Cyg, CIT~3, IK~Tau, $o$~Cet, and U~Ori), the spectral shapes
showed convincing agreement.  For example, U~Ori showed a relatively
flat silicate feature after one luminosity minimum, developing a
peakier feature at maximum.  Further data taken after the following
minimum showed again a flattened dust spectrum, while a final observation
taken at $\phi=0.01$ revealed a peaked spectrum similar to that
observed one cycle earlier.

This effect appears to have been first observed by Forrest, Gillett, \&
Stein (1975, hereafter FGS75), but was not connected to the
pulsational phase.  While a total of six $o$~Cet mid-IR spectra
spanning a 6~year period were presented, only the first observation occurred
sufficiently near maximum to show silicate enhancement.  The other
observations all showed significantly less silicate emission, leading
FGS75 to speculate that the infrared excess around this star was
decreasing slowly over time.  The data presented here clearly favors an
interpretation which links the amount and character of the silicate
emission to the phase of $o$~Cet.

The evolution of the silicate enhancement from luminosity minimum to
maximum could not be continuously followed for a few sources (R~Aqr and
VX~Sgr) due to inadequate temporal coverage. In these cases it is
possible that the silicate enhancement may have occurred suddenly,
and/or may not repeat from one cycle to the next.  This is especially
pronounced for VX~Sgr which shows a strong, narrow enhancement near
maximum ($\phi=0.84$). The most recent measurement, $\phi=0.85$ on
1997 August~29, may be starting to show a 9.7\,$\micron$ enhancement, but
further data at the end of 1997 is necessary to determine if this
feature re-appears and can indeed be associated with the pulsational cycle.

In order to investigate the physical mechanisms responsible for the
spectral changes, further analysis of the spectral shapes was performed
on a subset of the stars with the most complete temporal coverage.  On
this basis, the spectra of IK~Tau, U~Ori, and $o$~Cet were chosen to
represent the entire data set in subsequent analysis.  CIT~3 was
excluded because its mid-IR spectrum suggests that the dust shell is
optically thick.  These stars show the silicate feature enhancement
quite strongly and it is expected that the physical mechanism or
mechanisms responsible for the spectral changes in these sources should
apply to the others.

Subtracting the ``stellar component'' from the spectrum allows the
changes in the dust spectrum to be seen more clearly.  The stellar
spectrum is approximated as a 2500\,K blackbody and is fitted to the
8\,$\micron$ datum in the mid-IR spectrum (Little-Marenin \& Little
1990).  The ``stellar-subtracted'' spectra for IK~Tau, U~Ori, and
$o$~Cet are found in Figure~4.  All three sources show a relative
increase in the 9.7\,$\micron$ feature relative to the dust emission 
in the wings (near 12\,$\micron$, for example) which appears near the
maximum.   This is clear just by visual inspection, since the dust
emission at 12\,$\micron$ changes usually by less than a factor of 2,
while the peak flux varies by more than a factor of 2.
In addition, U~Ori shows a spectrally narrow ($\sim$0.5\,$\micron$)
feature near the silicate peak arising near maximum light.
While this feature is only apparent for U~Ori in Figure~4, 
the narrow character of the $\chi$~Cyg and R~Aqr silicate emission near
maximum (seen in Figure~2) may be of a similar nature.

\subsubsection{Appearance of ``Pure'' Silicates near Maximum
Luminosity} 
While the relatively large increase in 8-11\,$\micron$ flux as compared
to 11-13\,$\micron$ flux can be partially attributed to the dust shell
temperature increase during maximum light, we will show that the large
magnitude of the observed changes implies that the dust optical
constants themselves are changing.  Furthermore, the narrow emission
feature of U~Ori is clearly of non-thermal origin; the Planck function
has no such narrow features.  This is strong evidence that much of the
mid-IR emission near maximum light comes from relatively pure silicate
dust particles, as judged by the general narrowing of the silicate peak
and, in the case of U~Ori, by the extreme sharpness of the
9.7\,$\micron$ resonance.  Such pure silicates may be formed from
pre-existing, ``dirty'' grains which are ``purified'' via grain surface
evaporation of metal impurities or re-crystallization of amorphous
grains during the temperature rise towards maximum light (Tielens 1990;
Stencel et al.  1990; Dominik, Sedlmayr, \& Gail 1993).  Alternatively,
the dirty silicates formed at minimum light may be destroyed leaving
only small dust grains to emit their distinctive spectrum
(Little-Marenin et al.  1996).  The lack of the narrow U~Ori feature in
the IK~Tau and $o$~Cet spectra could be due to different shell
geometries, different chemical abundances, or peculiarities of 
recent pulsational cycles.

\subsubsection{Constructing the Color-Color Diagram}
Although the qualitative changes in the spectral shapes are plain to
see, it is useful to connect with previous studies of the silicate
emission feature.  A recent classification scheme which is easy to
implement yet reproduces many of the important features of earlier work
is given by Sloan \& Price (1995, hereafter SP95).  SP95 construct a
color-color diagram to define a one-parameter family of the mid-IR
spectral shapes.  Following Creech-Eakman et al. (1997), the SP95
procedure was further simplified by using a 2500\,K Planck function as
an approximate stellar spectrum and fitting it to the flux at
8\,$\micron$ (as described above).  Figure~5 was produced from the data
found in Figure~4.

SP95 define the ``spectral emission index,'' which varies from SE1 to
SE8, to parameterize the solid curve, beginning at
$\frac{F_{10}}{F_{11}}=.5$ to $\frac{F_{10}}{F_{11}}=1.4$ (see Figure~5
of SP95); the dividing lines are located on Figure~5.  The optical
constants of Draine \& Lee (1984) and Ossenkopf, Henning, \& Mathis
(1992), two dust optical property models commonly used in radiative
transfer calculations, describe the 9.7\,$\micron$ spectral signature
of amorphous silicate dust, a signature which places these spectra with
spectral emission indices SE5 to SE8.  SP95 further identify spectra
classified as SE4 to SE6 with spectra containing 10 and 11\,$\micron$
emission components, thought to be produced by warm crystalline olivine
(Little-Marenin \& Little 1990; Sloan \& Price 1995).  Although not
relevant to the spectra in this paper, the lower-left portion of the
SP95 color-color diagram is reserved for mid-infrared spectra which are
very broad (SE1 to SE3), thought to originate from alumina grains
(Onaka, de Jong, \& Willems 1989; Sloan \& Price 1995).  Because of the
difficulty in matching the data in Figure~5 with the corresponding
pulsational phases, we have plotted the spectral emission index as a
function of pulsational phase in Figure~6.

The data from IK~Tau, U~Ori, and $o$~Cet can be seen to span a
continuous range on the SP95 color-color diagram (see Figure 5),
starting at SE4 and ending with SE7.  While useful for classifying the
general shapes of silicate features, one can not expect the SP95
color-color diagram to fully describe the mid-infrared spectral shape
since only 3~discrete wavelengths are sampled, i.e. the narrow U~Ori
emission feature observed in this study would not have a strong
signature in Figure~5.  Figure~6 makes it clear that the silicate
feature variations are in phase with the pulsational cycle, showing
high spectral emission index values near maximum light and
significantly lower values just after minimum.

A variety of effects can cause changes in the spectral emission index
as a function of stellar luminosity.  Variations in the optical depth
or temperature of the dust could cause each star's location on the SP95
color-color diagram to shift, especially if self-absorption is
occurring at the peak of the silicate feature.  The phase-dependent
evolution of the dust optical constants responsible for changes in the
SP95 spectral emission index in Figures~5 and~6 may be related to the
narrow emission feature seen for U~Ori.  The following sections explore
these possibilities in more detail.

\subsubsection{Dust Self-Absorption}
If the temperature profile and optical properties of the dust shell
remain unchanged, then overall changes in the optical depth will not
cause a shift in the SP95 spectral emission index if the dust envelope
remains optically thin.  The Infrared Spatial Interferometer (ISI) has
obtained mid-IR visibility data for IK~Tau, $o$~Cet, and U~Ori, and
were fit, along with IRAS-LRS spectra, by circumstellar dust shell
models assuming a spherically symmetric outflow (Danchi et al. 1994).
These models indicate 11\,$\micron$ optical depths of $\sim$1.7,
$\sim$0.1, and $\sim$0.1 for IK~Tau, $o$~Cet, and U~Ori respectively.
Hence, the optically thin condition is expected to hold for $o$~Cet and
U~Ori, but not for IK~Tau.  This explains the more rounded nature of
silicate feature for IK~Tau and its lower spectral emission index,
consistent with self-absorption at the peak of the silicate feature.
$\chi$~Cyg, R~Aqr, and VX~Sgr were also observed and modeled by the
ISI, resulting in estimated 11\,$\micron$ optical depths of 0.43, 0.23,
and 0.56 respectively.  Hence, optical depth effects are not expected
to explain the spectral shape changes observed in most of the stars in
Figure~2 which show silicate feature enhancement near maximum light.

One would think that optical depth effects would be important for CIT~3
(see Figure~2), since the silicate feature for this star appears to be
self-absorbed, a sign that the optical depth near the peak is greater
than unity.  However, CIT~3 shows the same spectral changes as the
optically thin sources,  i.e. a relative enhancement near
9.7\,$\micron$ during maximum.  One could explain the spectral changes
with the presence of additional dust condensing near minimum light
causing the optical depth to increase, hence increasing the
self-absorption.  This is indeed a reasonable interpretation, however
such a scenario is not plausible for the optically thin sources in our
sample.  Furthermore, NML~Cyg has a mid-IR spectrum similar to CIT~3,
but shows no variation whatsoever (see Figure~1).  How can we explain
this difference?  This can be explained if CIT~3 is surrounded by a
detached shell which is causing the silicate absorption feature.  One
can then hypothesize an inner dust shell, similar to IK~Tau, internal
to an optically thick ``outer'' shell, which causes the spectral line
inversion.  NML~Cyg, on the other hand, has been recently modeled with
an inner dust shell radius at approximately 12 stellar radii (Monnier
et al. 1997).  This distant dust will probably not experience grain
formation and destruction during a pulsational cycle, since the grain
temperature is less than 1000\,K and gas densities are significantly
lower.  Additional mid-infrared and near-infrared interferometric
observations could help resolve this issue.

\subsubsection{Dust Shell Heating and Cooling}
One other physical mechanism for causing a change in spectral shape as
a function of luminosity is a change in the temperature.  While the
mid-infrared stellar spectral shape is relatively insensitive to changes in
stellar effective temperature (Little-Marenin \& Little 1990), the dust
spectrum depends strongly on dust shell temperature because the peak of
the Planck function lies within the mid-infrared for typical dust
temperatures, 500-1300\,K.  For optically thin dust envelopes following
a radial, power-law density distribution, the radial temperature
profile also follows a power-law relation from radiative transfer
calculations.  Hence, assuming no evolution of the optical constants
themselves,  changes in the spectral shape can be caused only
by changes in the dust temperature at the dust shell's inner radius,
not by changes in the optical depth or location of dust
condensation, from self-similarity arguments (for recent discussion,
see Ivezic \& Elitzur 1997).  Since the dust density distributions may
be shell-like (from episodic mass ejection), consideration is also
given to isothermal dust shell models.  The effect of dust shell
temperature variations on the spectral shape, as parameterized by the
SP95 color-color diagram, can be significant.  Figure~5 contains a
curve which shows the effect on the spectral emission index for an
isothermal shell at various temperatures (calculations for uniform outflow
models show roughly the same magnitude of variation and are not
included on the figure).  Although a shell which evolves from 500\,K
at minimum to 1300\,K at maximum almost matches the observed
variations, such large changes are implausible from radiative
transfer calculations and observations (typically $\Delta T \simle
$150\,K, Danchi et al. 1994).  

It is quite plausible then that half of the spectral changes observed
could be due to changes in the dust temperature, especially for cooler
dust shells where the emission spectrum is more sensitive to temperature.
However, reasonable physical constraints on the effective dust temperature and
the plausible temperature variation exclude dust heating and cooling from
being the sole explanation for the observed changes.
This is especially true for U~Ori, where the variations of the
spectral shape of U~Ori, especially near the peak, can be seen (Figure 4) to be
too narrow to be caused by changes in the temperature alone.

\subsubsection{Changing Dust Properties}
Finally, the observed variability in the spectral emission index could
be caused by changes in the dust optical constants themselves.  This
explanation has already been invoked to understand the narrow
9.7\,$\micron$ emission feature of U~Ori, but may explain the overall shape
changes as well.  Dust grain absorption characteristics may be strongly
dependent on the density and temperature of the ambient gas during
grain formation.  The adsorption of metal impurities onto the surface
of grains as well as the development of amorphous dust grains during
grain growth can have significant effects on the spectrum (cf.
Tielens 1990; Stencel et al. 1990; Dominik et al. 1993).
It is also known that grain size can affect the mid-IR emission spectra
of warm dust grains, but this requires grains with radii larger than a
micron (Papoular \& Pegourie 1983).  Such large grain sizes are
typically not considered in preparing radiative transfer models, where
conventionally the standard MRN dust grain size distribution
($a<0.25\,\micron$) observed in the ISM is utilized (Mathis, Rumpl, \&
Nordsieck 1977).

Theoretical studies of the nucleation of carbon grains (Fleischer,
Gauger, \& Sedlmayr 1992) indicate that dust formation is sensitively
dependent on density and temperature, thus luminosity fluctuations or
shock wave propagation can have unexpectedly large effects.  Silicate
grain formation theory is less well understood but efforts are
presently underway to create a self-consistent, spherically symmetric
model of the circumstellar environment of oxygen-rich miras,
incorporating full radiative transfer, pulsation, and dust formation
(Jeong et al. 1997).

\subsubsection{Summary}
In summary, a combination of dust shell temperature change and
pulsational phase-dependent evolution of the dust grain characteristics
can explain the observed spectral shape changes seen in Figure~2, while
the production of especially ``pure'' silicate grains is likely
required to explain the narrow ($\Delta \lambda \approx 0.5\,\micron$)
emission feature near the silicate peak during maximum light of a few
stars.  The general narrowing of the silicate peak, common to most of
the stars in Figure~2, may be due to the evaporation of impurities from
the grain surfaces or restructuring of the crystal lattice during
re-heating preceding maximum light.  Following maximum light,
impurities adsorb onto the grains, broadening the solid-state
resonance.  More detailed laboratory measurements of
astrophysically-realistic dust grains under varying physical conditions
would be necessary to place the above scenario on firm physical
footing.  While the optical depth is expected to increase near minimum
light as additional condensation occurs in the cooler circumstellar
environment, this by itself should not strongly effect the shape of the
dust spectrum if the dust optical properties remain fixed and
the envelope remains optically thin, a condition met
by most of the observed sources according to interferometric and
spectral measurements.

\subsection{Spectral Slope Changes or Enhanced Variability}
Figure~3 presents the spectral monitoring of stars whose spectral shape
showed either significant changes in slope or an rms shape variability
larger than that expected from observations of standard stars.  R~Cnc
and W~Aql showed evidence of a shallower spectral slope during minimum
luminosity than during maximum luminosity.  Such a change in slope
would be expected from heating and cooling of the dust shell.  Indeed,
simple calculations demonstrate that the observed changes are
consistent with $\Delta T_{\rm{dust}} \approx 100-200$\,K during the
luminosity cycle and $T_{\rm{dust}} \approx 500-800$\,K.  A more
detailed analysis can be done, but would require information regarding
the geometry of the dust shell, e.g. the location of dust shell inner
radius and dust density distribution.  

IRC~+10216 and R~Leo are included in Figure~3 due to their large rms
spectral shape fluctuations (IRC~+10216 also shows weak evidence for a
spectral slope change near maximum light).  Such fluctuations
may be indicative of a dynamic dust condensation zone, a zone permeated
with atmospheric shocks and subject to chaotic dust
formation/destruction processes.  Alternatively, these fluctuations may
be simply due to a relatively poorer calibration for these sources than
the rest of the data sample.  CIT~6 and  V~Hya (both carbon stars) also
showed rms fluctuations somewhat above the average of $<$$\sigma$$>
=1.3\%$, but these were not included here because they did not pass the
$\sigma>2\%$ cut.  Continued monitoring will be required to determine
the true source of the observed changes.

%___________________________________CONCLUSIONS________________________
\section{Conclusions}

The mid-infrared spectra of 30~late-type stars have been monitored in
order to detect changes occurring on the pulsational time scale
(typically 1-2~years) of long period variables (LPVs).  Stars which
exhibited little or no bolometric variability (i.e. non-LPVs) generally
showed no change in their spectral shape in the range 8-13\,$\micron$.
Furthermore, most stars with no strong 9.7\,$\micron$ silicate feature,
including carbon stars and oxygen-rich miras with broad, weak silicate
features, showed no spectral shape change.  However, a few such stars
in this category displayed either enhanced variability as a function of
wavelength (IRC~+10216 and R~Leo) or a detectable change in the
spectral slope correlated with pulsational phase (R~Cnc and W~Aql).
The former effect has no clear explanation, while the latter effect can
be explained by changes in the dust shell temperature ($\Delta
T_{\rm{dust}}\simle 200$\,K).

The most significant result presented here is that nearly all of the
observed sources with clear 9.7\,$\micron$ silicate features and
definite bolometric variability showed strong evidence for changes in
the silicate emission strength and spectral profile which are
correlated with pulsational phase.  We conclude that silicate
emission variation is a general property of long-period variables with
optically thin silicate features.  The sharpening of the silicate feature
near maximum light and its subsequent broadening can be explained by
the heating and cooling of the dust envelope coupled with
changing optical constants for the dust grains.  
The appearance of a
spectrally narrow emission feature near the silicate peak of a few
stars strongly indicates the existence of ``pure'' silicate dust
grains near maximum light.  We hypothesize that the general
narrowing of the dust emission
spectra may arise from pre-existing dirty grains whose surface
impurities have been evaporated off or whose amorphous molecular
configuration has crystallized during the dust re-heating following
minimum luminosity.  The solid-state resonance would naturally broaden
after maximum light as impurities re-adsorb onto the cooled grain
surface.  Such speculation awaits more detailed laboratory measurements
of astrophysically-relevant grain types.

The observations presented here remind us that the dust formation
process is still only partially understood.  Indeed, uncertainties in
the optical constants for circumstellar dust are a primary obstacle in
creating self-consistent multi-wavelength radiative transfer models
incorporating interferometric observations of dusty objects (e.g.,
Monnier et al. 1997).  The changes observed in silicate optical
properties as a function of pulsational phase are not predicted by any
present dust formation theories, and more careful consideration is
required of the effects of photospheric shocks propagating into the
dust formation zone and of the changing temperature and density
structure due to the pulsation.  Such models may then not only explain
the changing optical properties of the dust around a single, pulsating
object, but may also explain why different stars possess silicate
emission with distinctly different spectral profiles.

%___________________________________ACKNOWLEDGEMENTS___________________
\acknowledgments
{The authors would like to thank Charles Townes, Peter Tuthill, Chris
Matzner, and Irene Little-Marenin for valuable comments.  This
research has made extensive use of the SIMBAD database, operated at
CDS, Strasbourg, France.  These observations were made as part of a
long-term infrared stellar interferometry program at U.C. Berkeley,
supported by the National Science Foundation (Grants AST-9315485,
AST-9321289, \& AST-9500525) and by the Office of Naval Research (OCNR
N00014-89-J-1583  \& FDN0014-96-1-0737).  Some of these spectra were
obtained as part of the UKIRT Service Programme.  The United Kingdom
Infrared Telescope is operated by the Joint Astronomy Centre on behalf
of the U.K. Particle Physics and Research Council.

} \pagebreak
%___________________________________BIBLIOGRAPHY_______________________
%\begin{thebibliography}{} 

\clearpage
\begin{deluxetable}{lccl}
\tablewidth{0pt}
\tablecaption{Journal of UKIRT Spectrophotometry}
\tablehead{
\colhead{Date}  &   \colhead{Number of}  &
\colhead{Absolute Flux}  &
\colhead{Comments}        \\ 
\colhead{(U.T.)} & \colhead{Spectra} & \colhead{Calibration} & }
\startdata
1994 August 26 & 20 & $\pm 5$\%  & Cloudy for $\alpha$~Boo, RX~Boo, $\alpha$~Sco, 
VX~Sgr, $\delta$~Oph \\
1994 November 20 & 10 & $\pm 20$\% & Not photometric \\ 
1995 January 14  & 19 & $\pm 5$\%  &                 \\
1995 March 17 & 24 & $\pm 5$\%  & Poor calibration for 9.7\,$\micron$ ozone line 
and band edges \\
1995 August 20 & 18 & $\pm 10$\% &  \\
1995 December 2 & 15 & $\pm 10$\% &  \\
1996 June 22 & 6  & $\pm 15$\% & Patchy cirrus \\
1996 September 26 & 4  & $\pm 5$\%  &  \\
1997 June 9 & 3  & $\pm 10$\% & \\
1997 August 28 & 1  & $\pm 10$\% & \\
1997 August 29 & 5  & $\pm 5$\%  & VX Sgr absolute calibration is uncertain \\
\enddata
\end{deluxetable}

%%%%%%%%%%%%%%%%%%%%%%%%%%%%%%%%%%%%%%%%%%%%%%%%%%%%%%%%%%%%%%%%%%
% Table Two contains the Stellar characteristics                 %
%%%%%%%%%%%%%%%%%%%%%%%%%%%%%%%%%%%%%%%%%%%%%%%%%%%%%%%%%%%%%%%%%%

\begin{deluxetable}{lcccccc}  
%\singlespace
%\arraystretch{.5}
\tablewidth{0pt}
\tablecaption{Stellar Characteristics}
\tablehead{
\colhead{Names} & \colhead{Spectral} & \colhead{Variable} & \colhead{Date of} & \colhead{Period} & \colhead{\# of}    & \colhead{References} \\
		& \colhead{Type\tablenotemark{a}} & \colhead{Type$\tablenotemark{a}$} & \colhead{Maximum} & \colhead{(days)} & \colhead{Spectra} & 			 
}
\startdata
$\alpha$ Boo	& K1.5III  	     & 			  & 		      &			 & 4			  & 			 \\
$\alpha$ Her	& M5		     &			  &		      & 		 & 3			  & 			 \nl
$\alpha$ Ori	& M1.5Iab	     &	SRc		  &		      & $\sim$2070	 & 7			  & 1 			 \nl
$\alpha$ Sco	& M1.5Ib	     &  SRa		  &		      & $\sim$1733	 & 3			  & 1			 \nl
$\alpha$ Tau	& K5III		     &		 	  &		      &			 & 6			  & 			 \\
$\beta$ Peg	& M2.5II-III	     & 			  &		      &			 & 1			  &			 \\
$\chi$ Cyg	& S8,K0III	     &	Mira		  & 2448688.5	      & 402.3		 & 4			  & 2		   	 \\
CIT 3=WX~Psc=IRC+10011& M9 &  Mira		  & 2446985	      & 660		 & 7			  & 5			 \\
CIT 6=RW LMi=IRC+30219& C		     &  Mira		  & 2449857	      & 628              & 4		          & 6			 \\
Egg Nebula=AFGL 2688& F5Iae		     &			  &	    	      &		         & 1 			  &			 \\
$\delta$ Oph	& M0.5III	     & 			  &		      &			 & 4 		          &			 \\
g Her		& M6III		     &  SRb		  &  		      & $\sim$70	 & 4		 	  & 1			 \\
IK Tau=IRC+10050& M6me		     &  Mira		  & 2447412	      & 462		 & 8			  & 5 			 \\
IRC+10216=CW Leo&  C & Mira		  & 2446825.0	      & 638		 & 5			  & 3,4			 \\
IRC+10420=V1302 Aql& F8Ia		     &			  &		      &			 & 6			  &  			 \\
NML Cyg=IRC+40448& {MI\tablenotemark{b}}  &  SR		  &		      & $\sim$940	 & 5			  & 7 			 \\
$o$~Cet=Mira    & M7IIIe	     &  Mira		  & 2448826.3	      & 333.8		 & 7			  & 2			 \\
R Aqr		& M7IIIpe	     &  Mira		  & 2448619.9         & 383.9		 & 3			  & 2			 \\
R Cnc		& M7IIIe	     &  Mira		  & 2448587.6	      & 356.0		 & 4			  & 2			 \\
R Hya		& M7IIIe	     &  Mira		  & 2448605.2	      & 384		 & 3			  & 2			 \\
R Leo		& M8IIIe	     &  Mira		  & 2448565.6	      & 311.0		 & 5			  & 2			 \\
RX Boo		& M7.5III	     &  SRb		  & 2449438		      & 340		 & 4			  & 9			 \\
SW Vir=IRC+00230& M7III		     &  SRb		  & 2448637.12	      & 153.6        	 & 2			  & 2			 \\
U Her		& M7III		     &  Mira		  & 2448652.8	      & 418		 & 1			  & 2			 \\
U Ori		& M8III		     &  Mira		  & 2448582.0	      & 367 		 & 5			  & 2			 \\
V Hya 		& C9		     &  SRa		  & 2447950		      & $\sim$529.2	 & 4			  & 1,9 	 \\
VX Sgr		& M4Iae		     &  SRb		  & 2448591	      & 737		 & 5			  & 2			 \\
VY CMa		& M5Iae		     &  SR		  &		      & $\sim$2000	 & 4			  & 8 			 \\
W Aql		& S4.9		     &  Mira		  & 2450762	      & 490		 & 3			  & 10,11\\
W Hya		& M8e		     &  SRa 		  & 2448790.0	      & 369		 & 3			  & 2			 \\
\enddata
\tablenotetext{a}{From Simbad Database}
\tablenotetext{b}{Supergiant classification based on Morris \& Jura (1983)}
\tablerefs{
1. Kukarkin et al. 1971; 2. ESA 1997 (Hipparcos Catalog); 
3. Le Bertre 1992; 4. Dyck et al. 1991;
5. Le Bertre 1993; 6. Alksnis 1995;
7. Monnier et al. 1997; 8. Marvel 1996;
9. Mattei 1995; 10. Vardya 1988;
11. Mattei 1997.
}
\end{deluxetable}

%%%%%%%%%%%%%%%%%%%%%%%%%%%%%%%%%%%%
%% Next goes the Figure Captions  %%
%%%%%%%%%%%%%%%%%%%%%%%%%%%%%%%%%%%%

\clearpage

\begin{figure}
\figurenum{1a}
%\plotone{figure1a.eps}
\plotfiddle{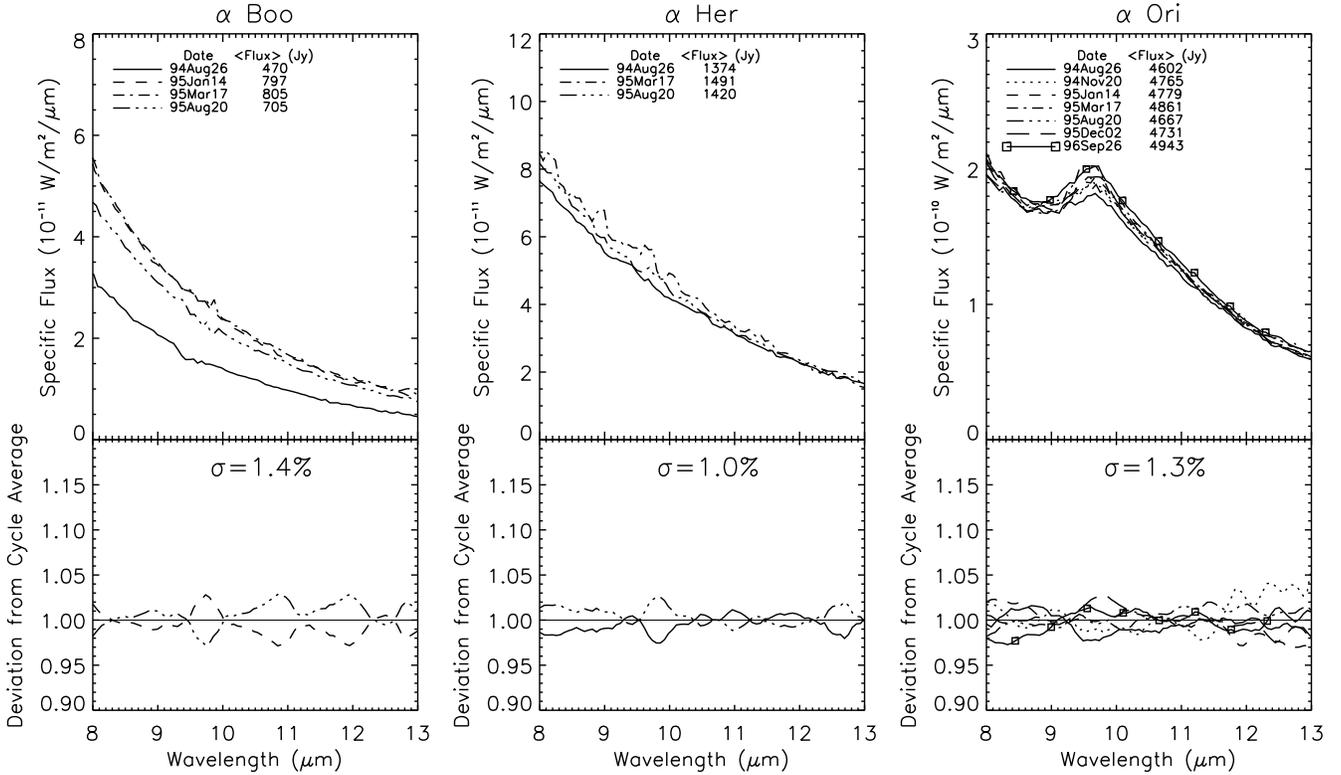}{4 in}{90.0}{75.0}{75.0}{305.0}{0}
\caption{Mid-infrared spectra of late-type stars which
showed no apparent changes in shape (see \S3.1).  The top panels show
the calibrated spectra along with the dates of observation, mean
(8-13\,$\micron$) fluxes in Janskys, and the pulsational phases, if
applicable.  All spectra taken on a given date share a common linestyle
to facilitate intercomparison.  The bottom panels show each spectrum's
deviation from the cycle-averaged spectral shape, with the rms spectral
shape deviation denoted by $\sigma$ (see \S3).}
\end{figure}

\begin{figure}
\figurenum{1b}
%\plotone{figure1b.eps}
\plotfiddle{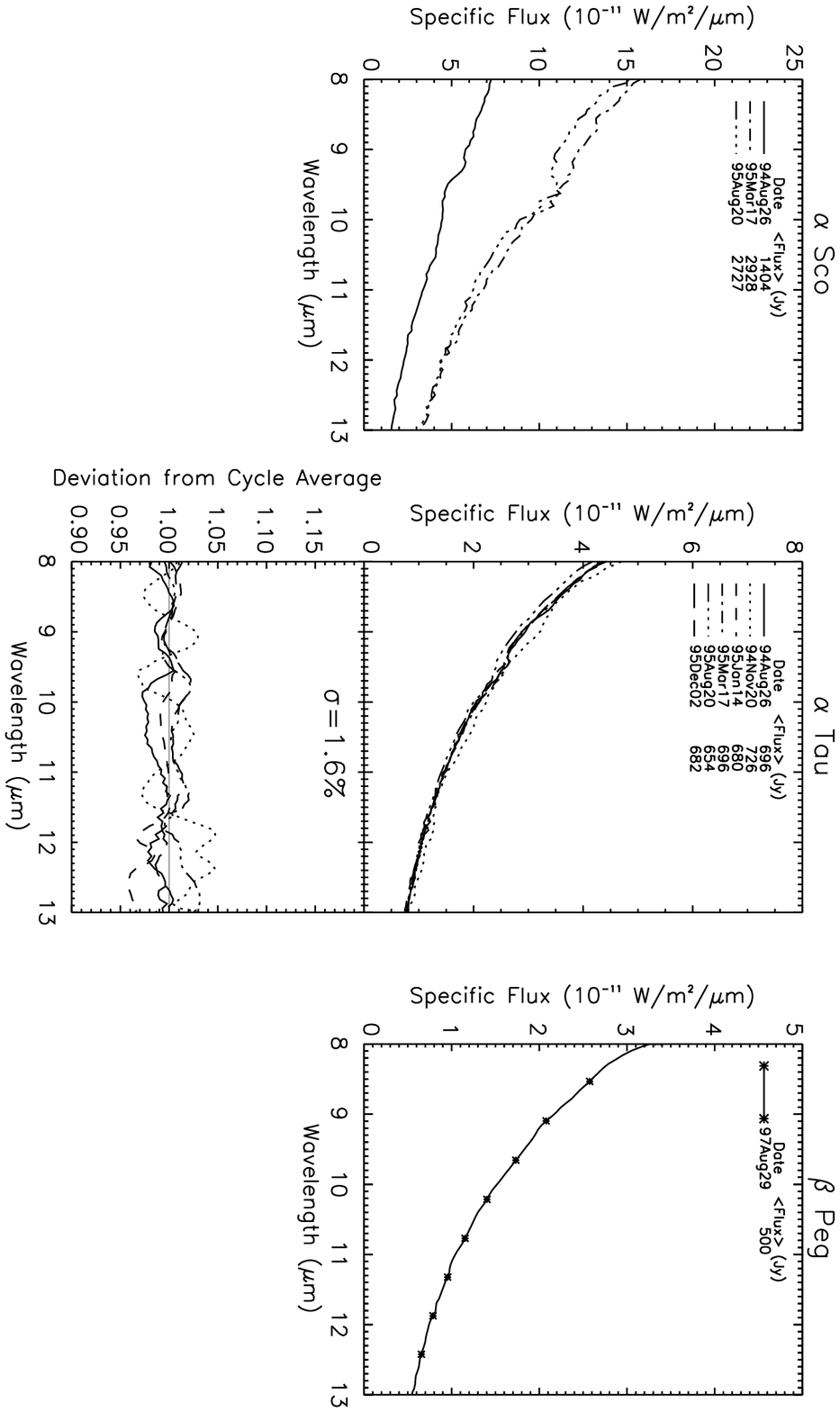}{4 in}{90.0}{75.0}{75.0}{305.0}{0}
\caption{Mid-infrared spectra of late-type stars which
showed no apparent changes in shape (see \S3.1).  The top panels show
the calibrated spectra along with the dates of observation, mean
(8-13\,$\micron$) fluxes in Janskys, and the pulsational phases, if
applicable.  All spectra taken on a given date share a common linestyle
to facilitate intercomparison.  The bottom panels show each spectrum's
deviation from the cycle-averaged spectral shape, with the rms spectral
shape deviation denoted by $\sigma$ (see \S3).}
\end{figure}

\begin{figure}
\figurenum{1c}
%\plotone{figure1c.eps}
\plotfiddle{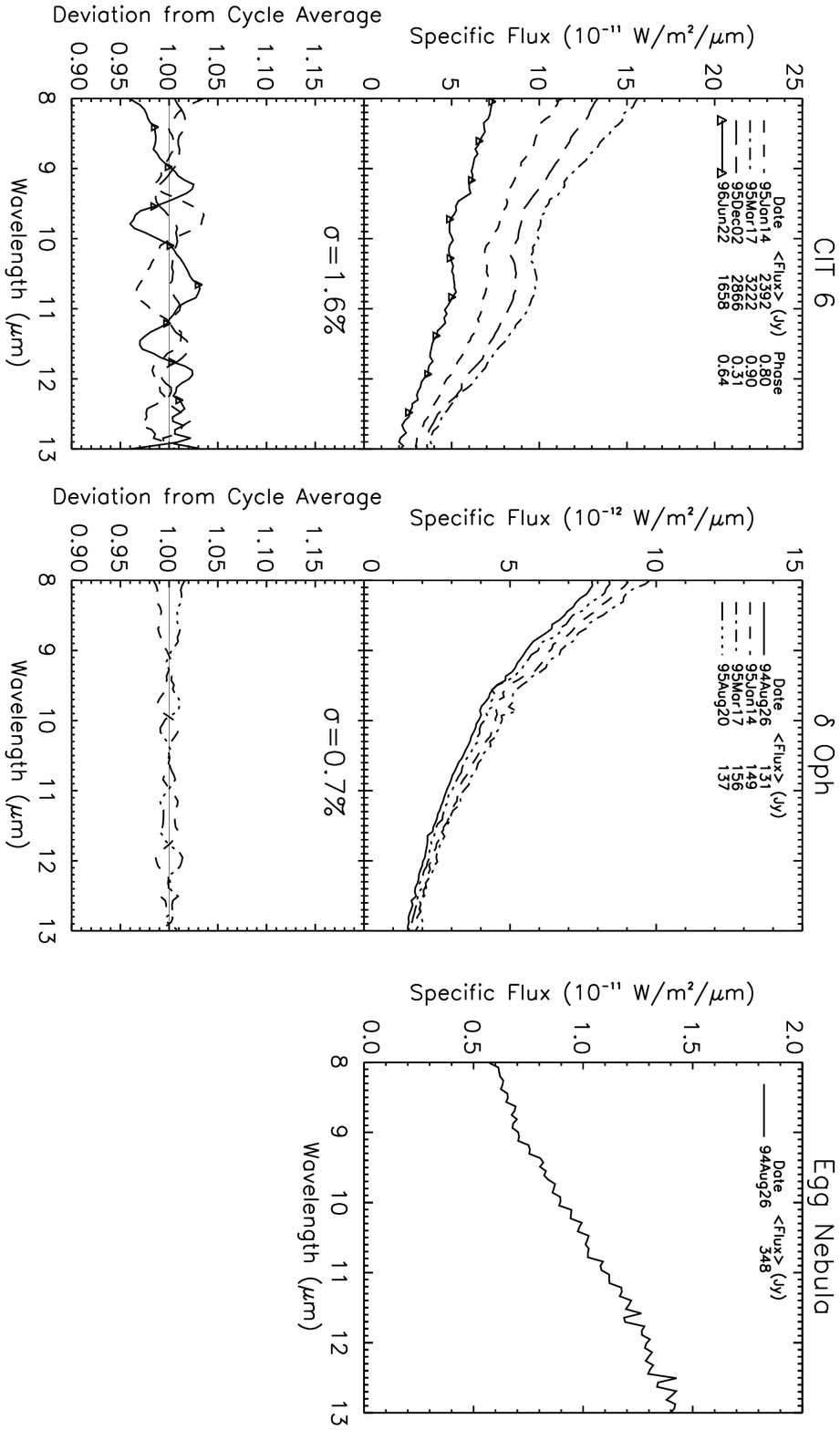}{4 in}{90.0}{75.0}{75.0}{305.0}{0}
\caption{Mid-infrared spectra of late-type stars which
showed no apparent changes in shape (see \S3.1).  The top panels show
the calibrated spectra along with the dates of observation, mean
(8-13\,$\micron$) fluxes in Janskys, and the pulsational phases, if
applicable.  All spectra taken on a given date share a common linestyle
to facilitate intercomparison.  The bottom panels show each spectrum's
deviation from the cycle-averaged spectral shape, with the rms spectral
shape deviation denoted by $\sigma$ (see \S3).}
\end{figure}

\begin{figure}
\figurenum{1d}
%\plotone{figure1d.eps}
\plotfiddle{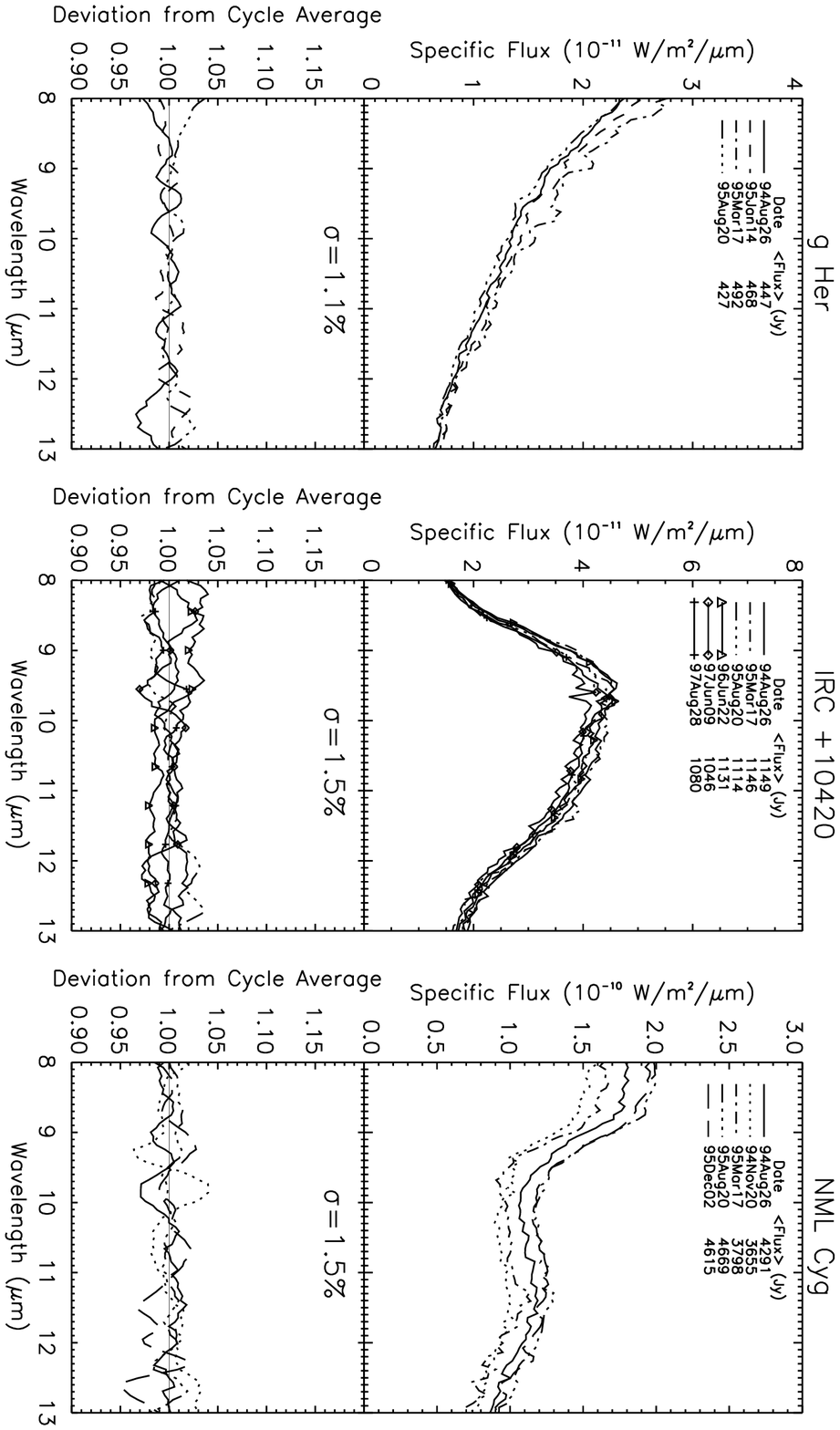}{4 in}{90.0}{75.0}{75.0}{305.0}{0}
\caption{Mid-infrared spectra of late-type stars which
showed no apparent changes in shape (see \S3.1).  The top panels show
the calibrated spectra along with the dates of observation, mean
(8-13\,$\micron$) fluxes in Janskys, and the pulsational phases, if
applicable.  All spectra taken on a given date share a common linestyle
to facilitate intercomparison.  The bottom panels show each spectrum's
deviation from the cycle-averaged spectral shape, with the rms spectral
shape deviation denoted by $\sigma$ (see \S3).}
\end{figure}

\begin{figure}
\figurenum{1e}
%\plotone{figure1e.eps}
\plotfiddle{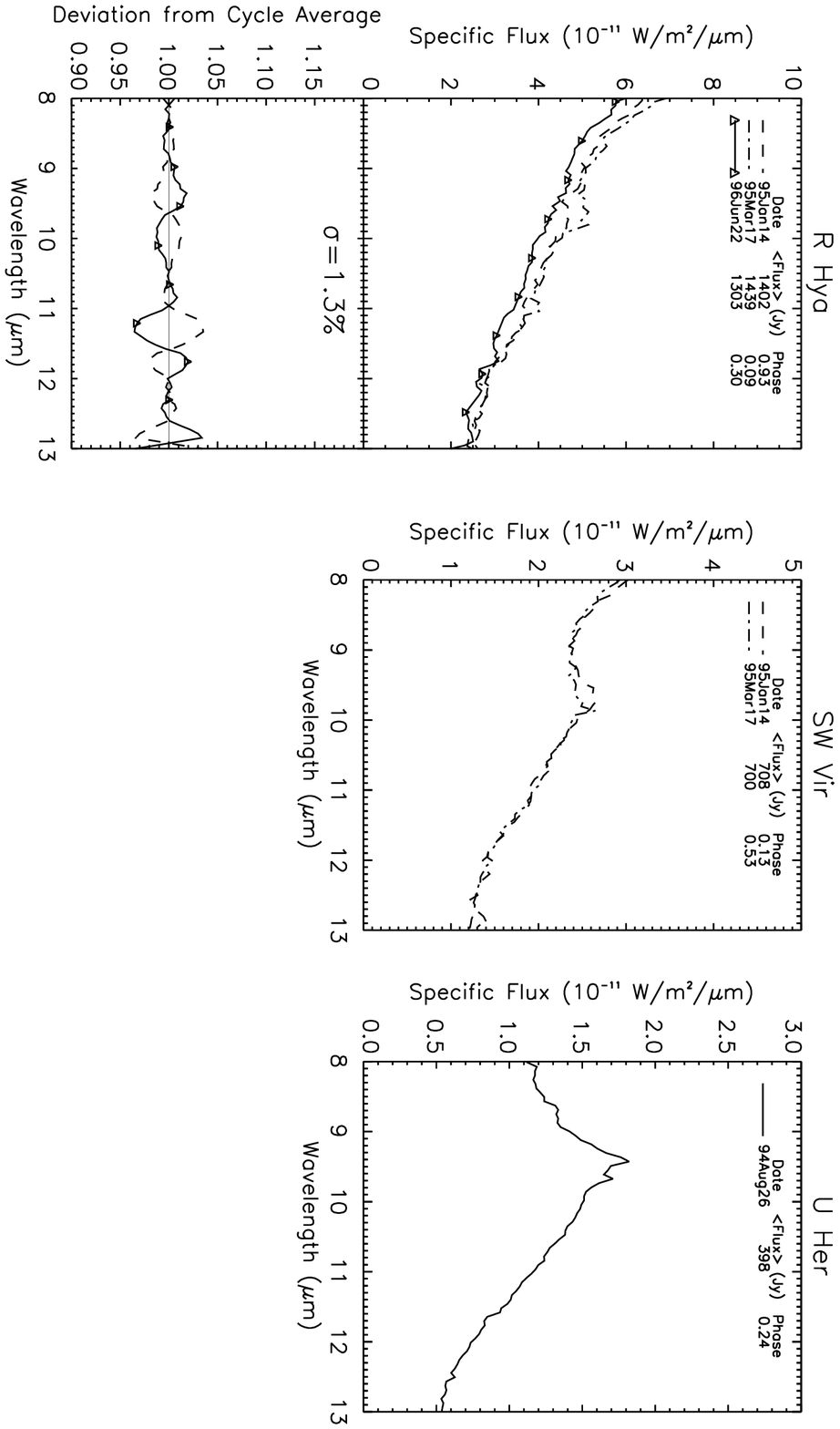}{4 in}{90.0}{75.0}{75.0}{305.0}{0}
\caption{Mid-infrared spectra of late-type stars which
showed no apparent changes in shape (see \S3.1).  The top panels show
the calibrated spectra along with the dates of observation, mean
(8-13\,$\micron$) fluxes in Janskys, and the pulsational phases, if
applicable.  All spectra taken on a given date share a common linestyle
to facilitate intercomparison.  The bottom panels show each spectrum's
deviation from the cycle-averaged spectral shape, with the rms spectral
shape deviation denoted by $\sigma$ (see \S3).}
\end{figure}

\begin{figure}
\figurenum{1f}
%\plotone{figure1f.eps}
\plotfiddle{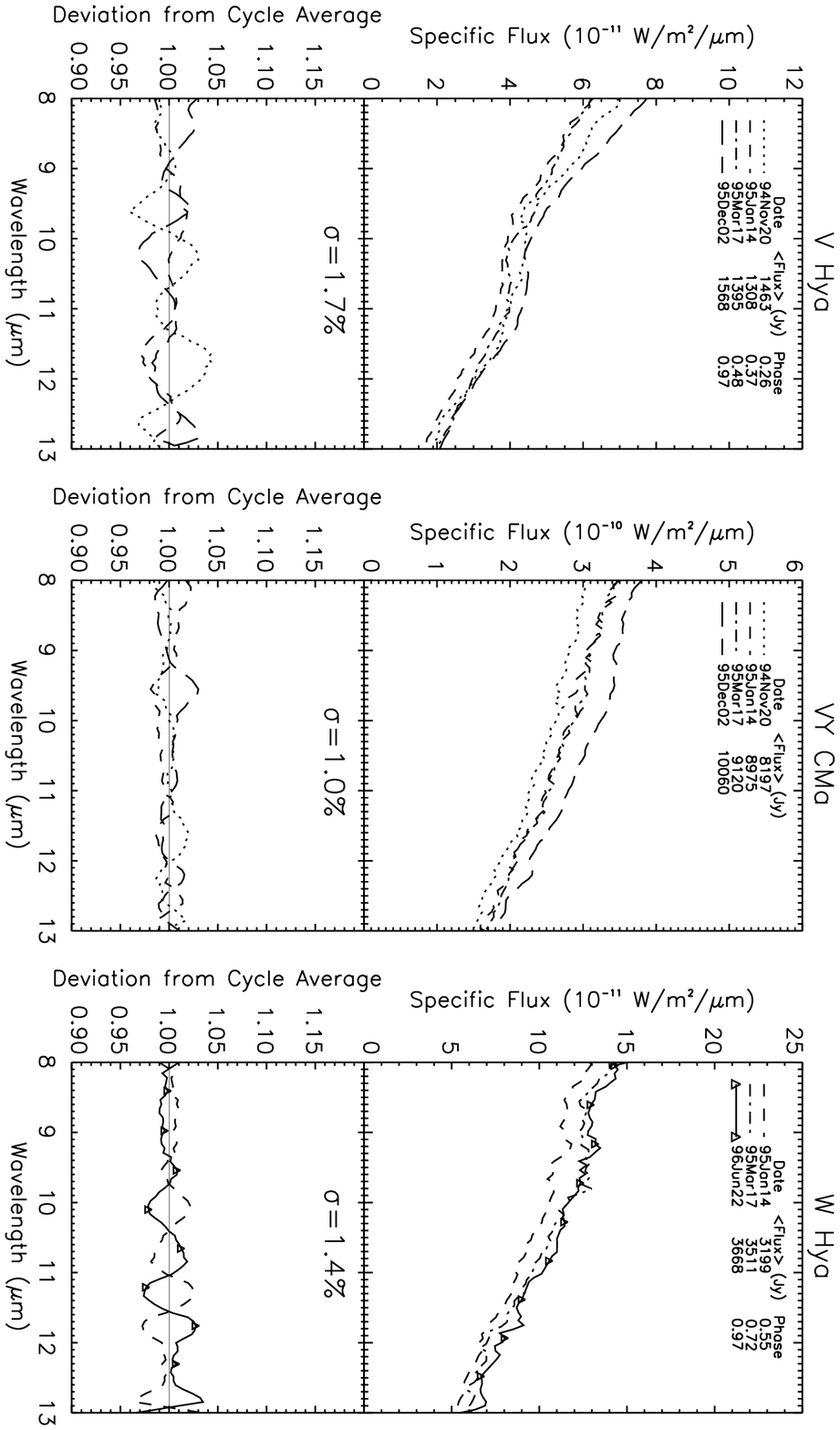}{4 in}{90.0}{75.0}{75.0}{305.0}{0}
\caption{Mid-infrared spectra of late-type stars which
showed no apparent changes in shape (see \S3.1).  The top panels show
the calibrated spectra along with the dates of observation, mean
(8-13\,$\micron$) fluxes in Janskys, and the pulsational phases, if
applicable.  All spectra taken on a given date share a common linestyle
to facilitate intercomparison.  The bottom panels show each spectrum's
deviation from the cycle-averaged spectral shape, with the rms spectral
shape deviation denoted by $\sigma$ (see \S3).}
\end{figure}

\begin{figure}
\figurenum{2a}
%\plotone{figure2a.eps}
\plotfiddle{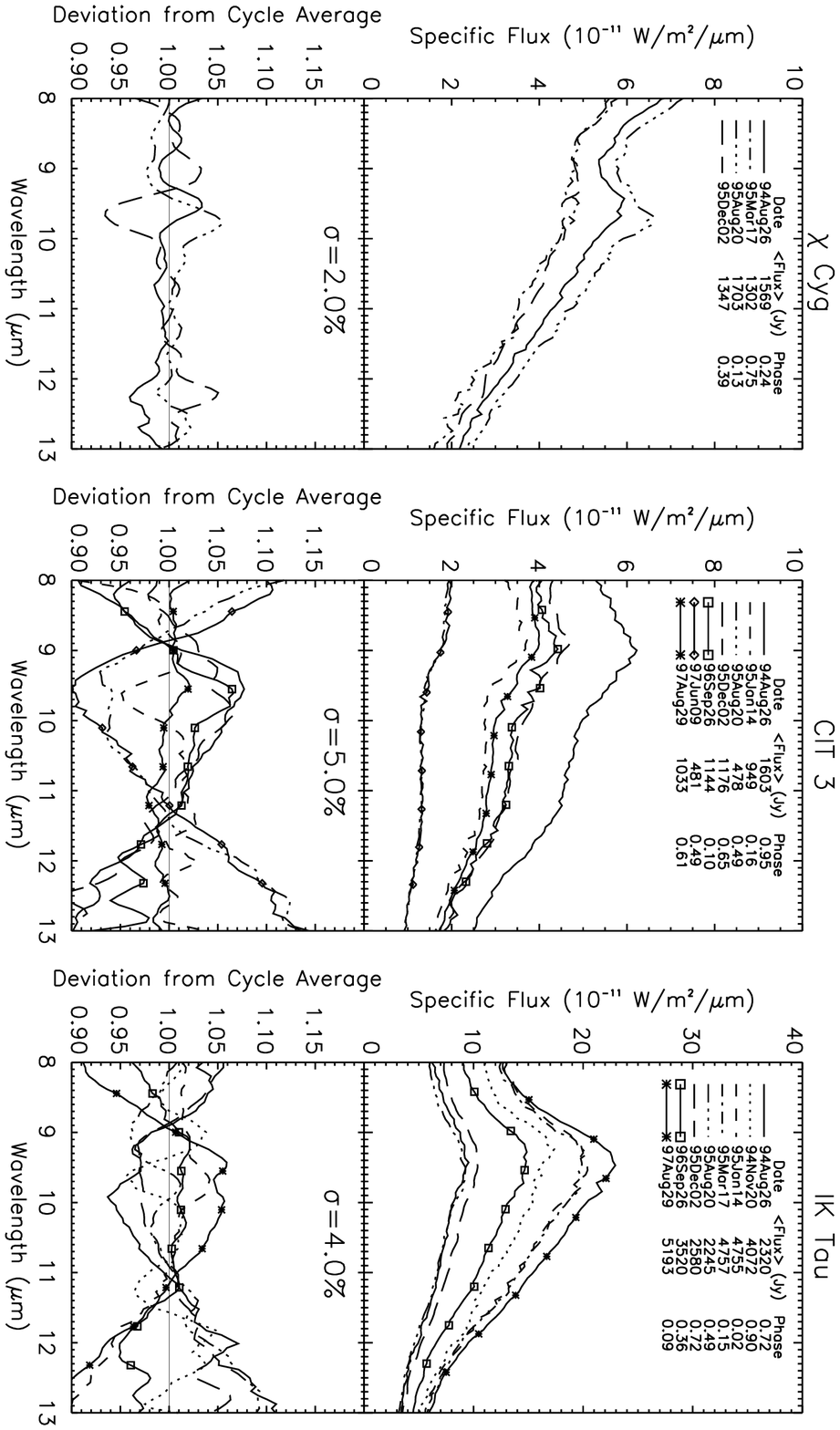}{4 in}{90.0}{75.0}{75.0}{305.0}{0}
\caption{Mid-infrared spectra of late-type stars whose
silicate feature showed significant enhancement during maximum light
(see \S3.2).  The top panels show the calibrated spectra along with the
dates of observation, mean (8-13\,$\micron$) fluxes in Janskys, and the
pulsational phases.  All spectra taken on a given date share a common
linestyle to facilitate intercomparison.  The bottom panels show each
spectrum's deviation from the cycle-averaged spectral shape, with the rms
spectral shape deviation denoted by $\sigma$ (see \S3).}
\end{figure}

\begin{figure}
\figurenum{2b}
%\plotone{figure2b.eps}
\plotfiddle{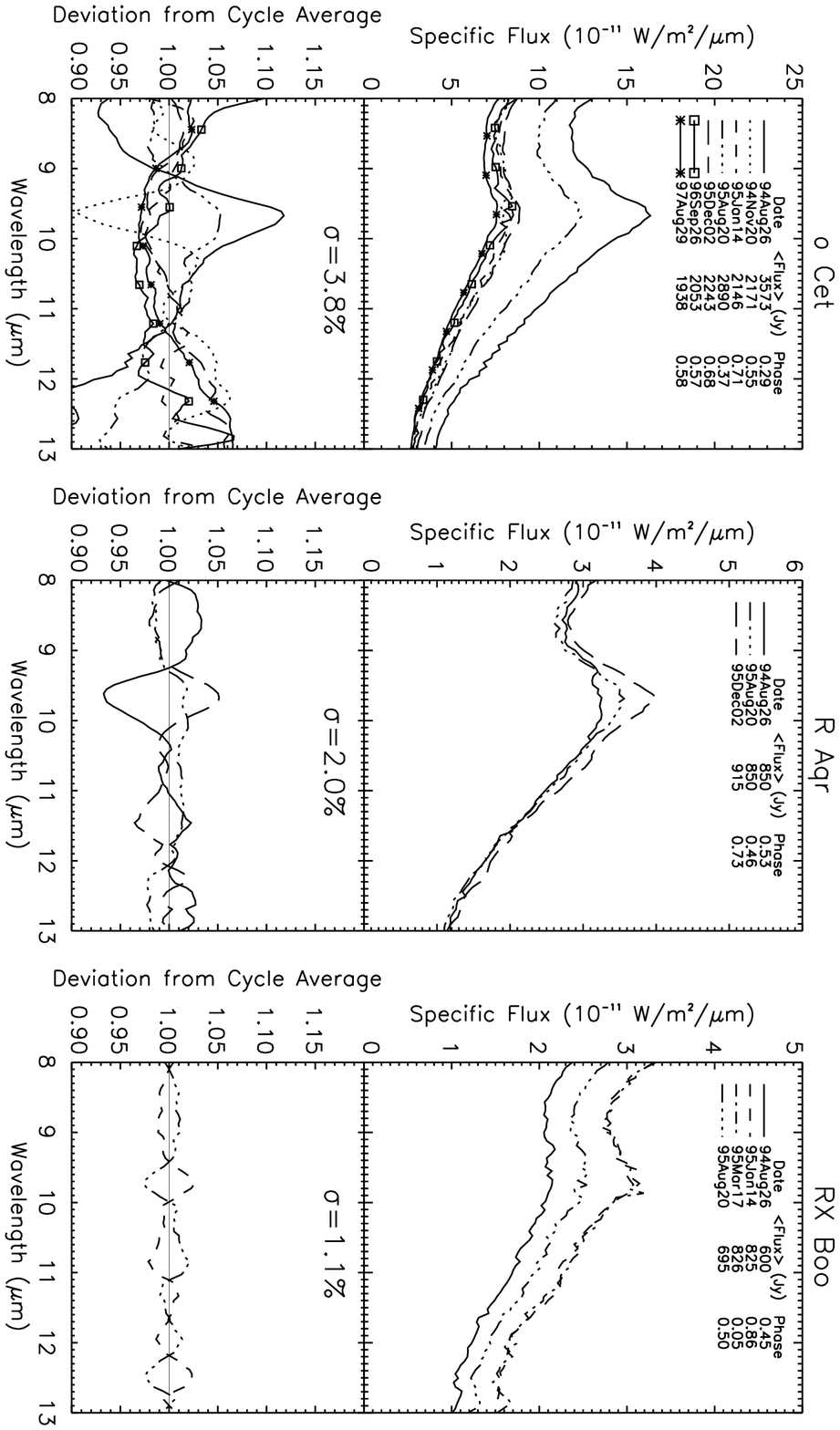}{4 in}{90.0}{75.0}{75.0}{305.0}{0}
\caption{Mid-infrared spectra of late-type stars whose
silicate feature showed significant enhancement during maximum light
(see \S3.2).  The top panels show the calibrated spectra along with the
dates of observation, mean (8-13\,$\micron$) fluxes in Janskys, and the
pulsational phases.  All spectra taken on a given date share a common
linestyle to facilitate intercomparison.  The bottom panels show each
spectrum's deviation from the cycle-averaged spectral shape, with the rms
spectral shape deviation denoted by $\sigma$ (see \S3).}
\end{figure}

\begin{figure}
\figurenum{2c}
%\plotone{figure2c.eps}
\plotfiddle{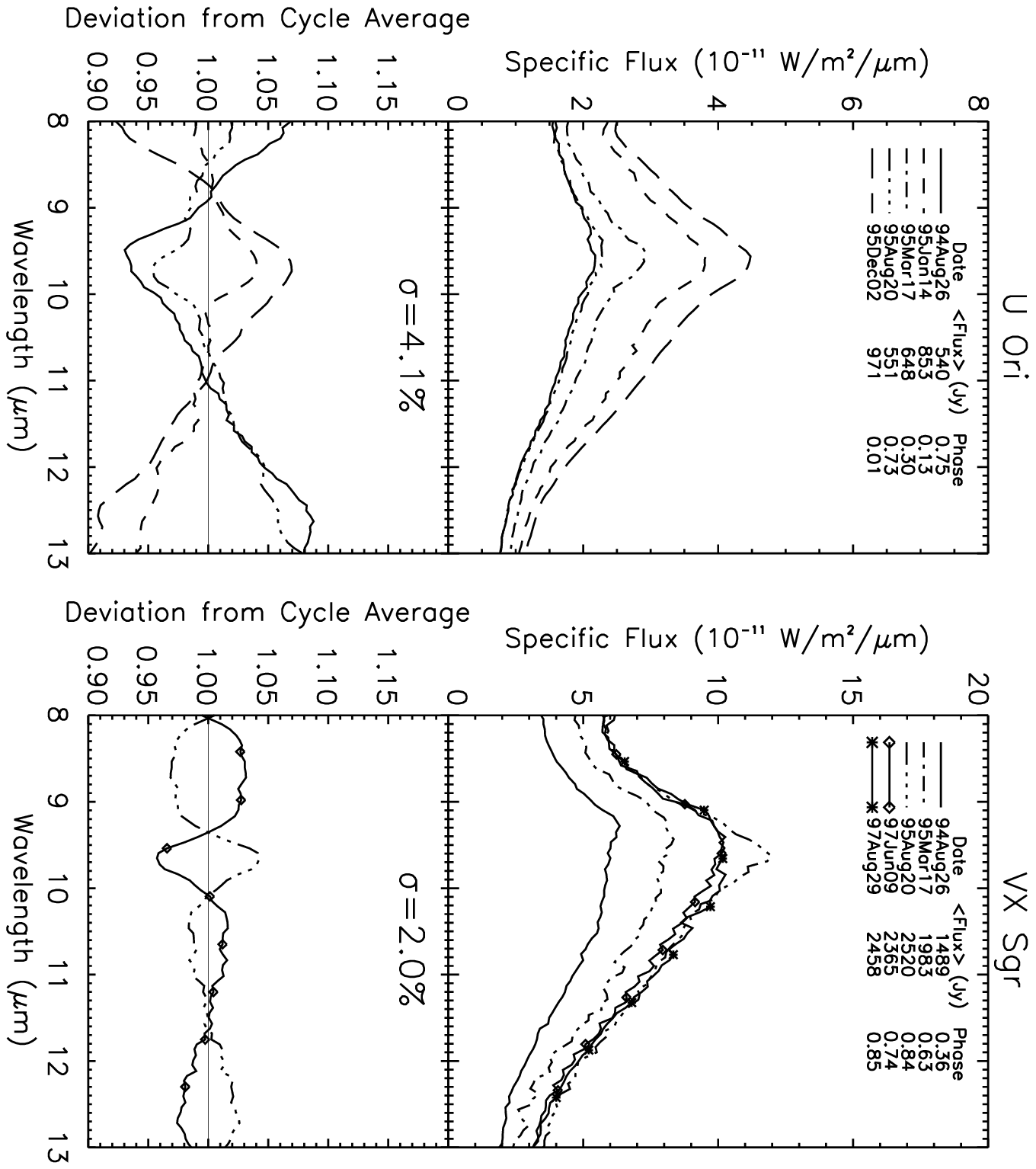}{4 in}{90.0}{75.0}{75.0}{385.0}{0}
\caption{Mid-infrared spectra of late-type stars whose
silicate feature showed significant enhancement during maximum light
(see \S3.2).  The top panels show the calibrated spectra along with the
dates of observation, mean (8-13\,$\micron$) fluxes in Janskys, and the
pulsational phases.  All spectra taken on a given date share a common
linestyle to facilitate intercomparison.  The bottom panels show each
spectrum's deviation from the cycle-averaged spectral shape, with the rms
spectral shape deviation denoted by $\sigma$ (see \S3).}
\end{figure}

\begin{figure}
\figurenum{3a}
%\plotone{figure3a.eps}
\plotfiddle{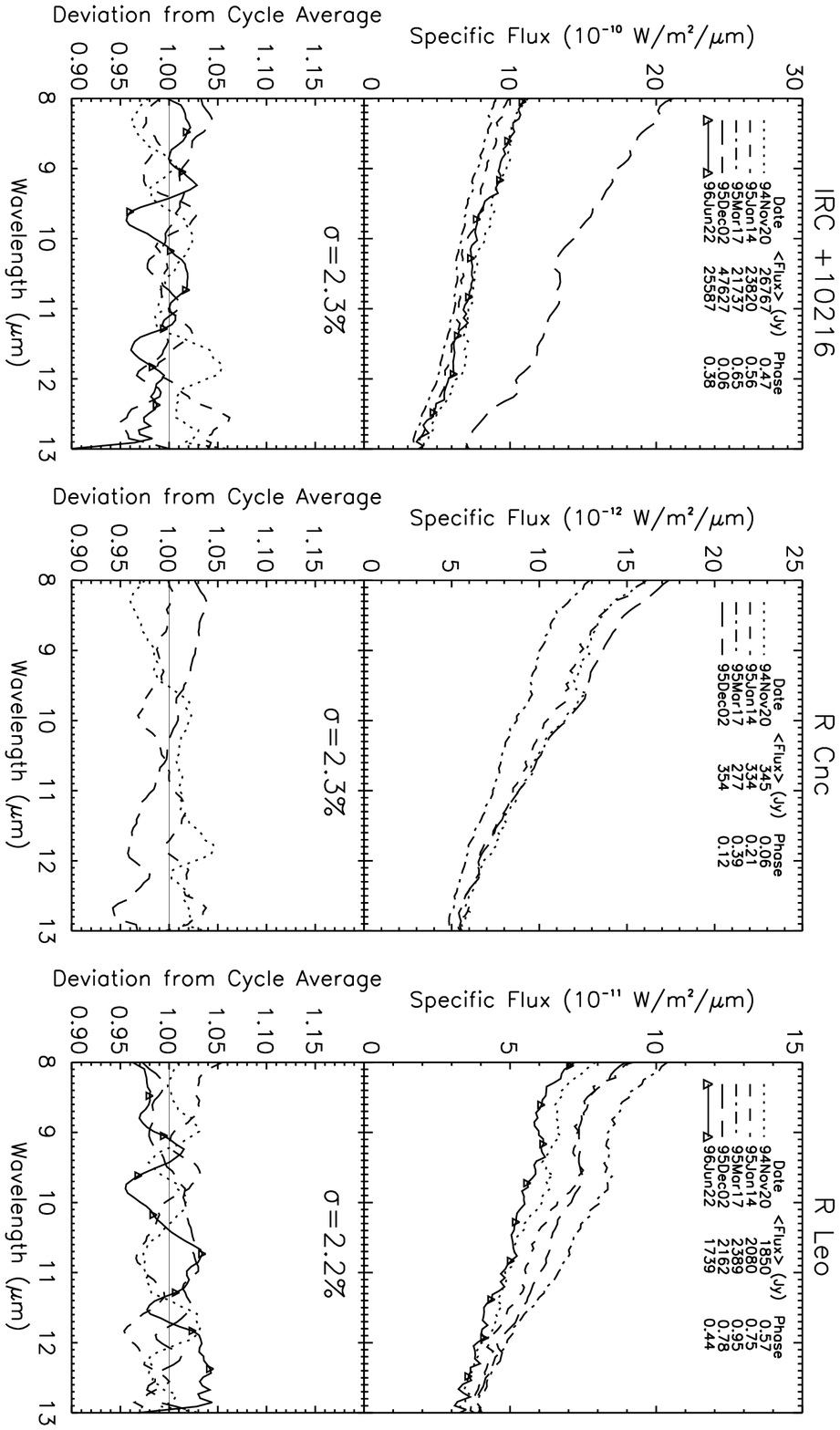}{4 in}{90.0}{75.0}{75.0}{305.0}{0}
\caption{Mid-infrared spectra of late-type stars which
showed either significant changes in slope or an rms variability larger
than that expected from observations of standard stars (see \S3.3).
The top panels show the calibrated spectra along with the dates of
observation, mean (8-13\,$\micron$) fluxes in Janskys, and the
pulsational phases.  All spectra taken on a given date share a common
linestyle to facilitate intercomparison.  The bottom panels show each
spectrum's deviation from the cycle-averaged spectral shape, with the rms
spectral shape deviation denoted by $\sigma$ (see \S3).}
\end{figure}

\begin{figure}
\figurenum{3b}
%\plotone{figure3b.eps}
\plotfiddle{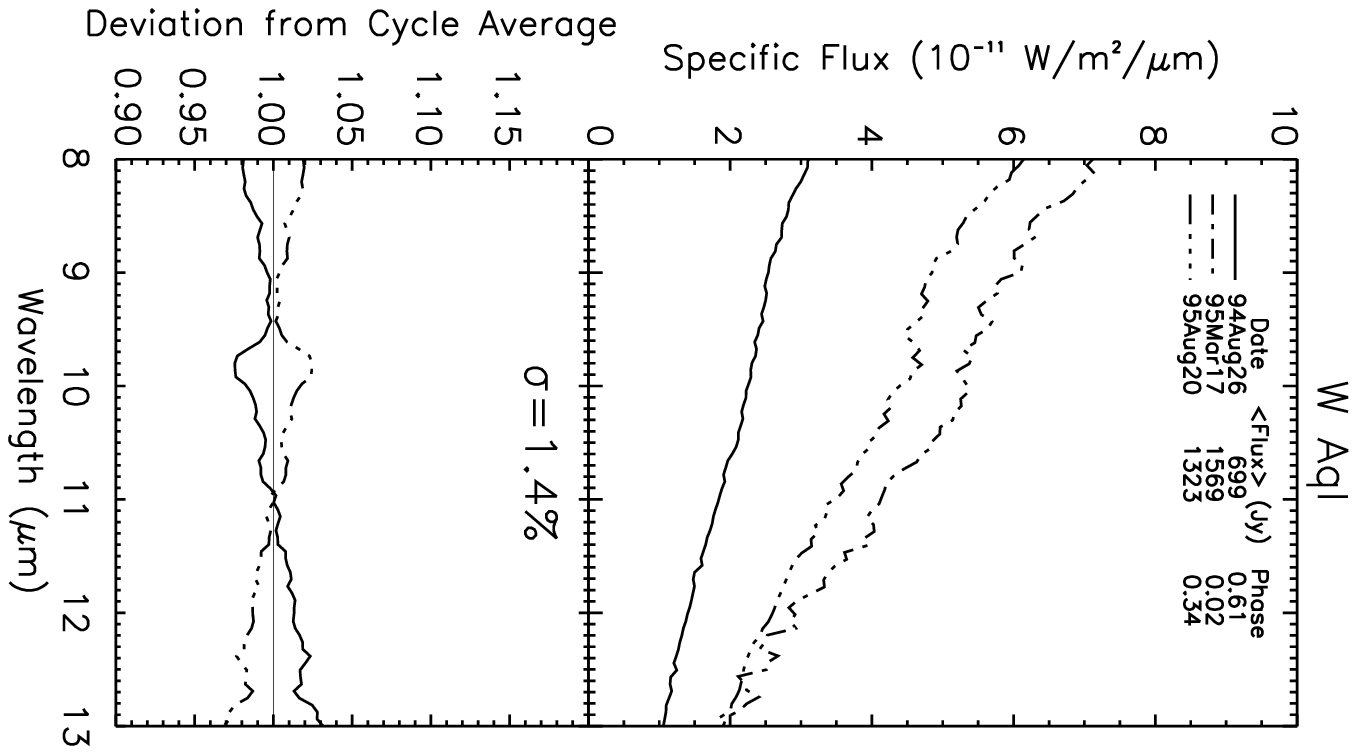}{4 in}{90.0}{75.0}{75.0}{465.0}{0}
\caption{Mid-infrared spectra of late-type stars which
showed either significant changes in slope or an rms variability larger
than that expected from observations of standard stars (see \S3.3).
The top panels show the calibrated spectra along with the dates of
observation, mean (8-13\,$\micron$) fluxes in Janskys, and the
pulsational phases.  All spectra taken on a given date share a common
linestyle to facilitate intercomparison.  The bottom panels show each
spectrum's deviation from the cycle-averaged spectral shape, with the rms
spectral shape deviation denoted by $\sigma$ (see \S3).}
\end{figure}

\begin{figure}
\figurenum{4}
%\plotone{figure4.eps}
\plotfiddle{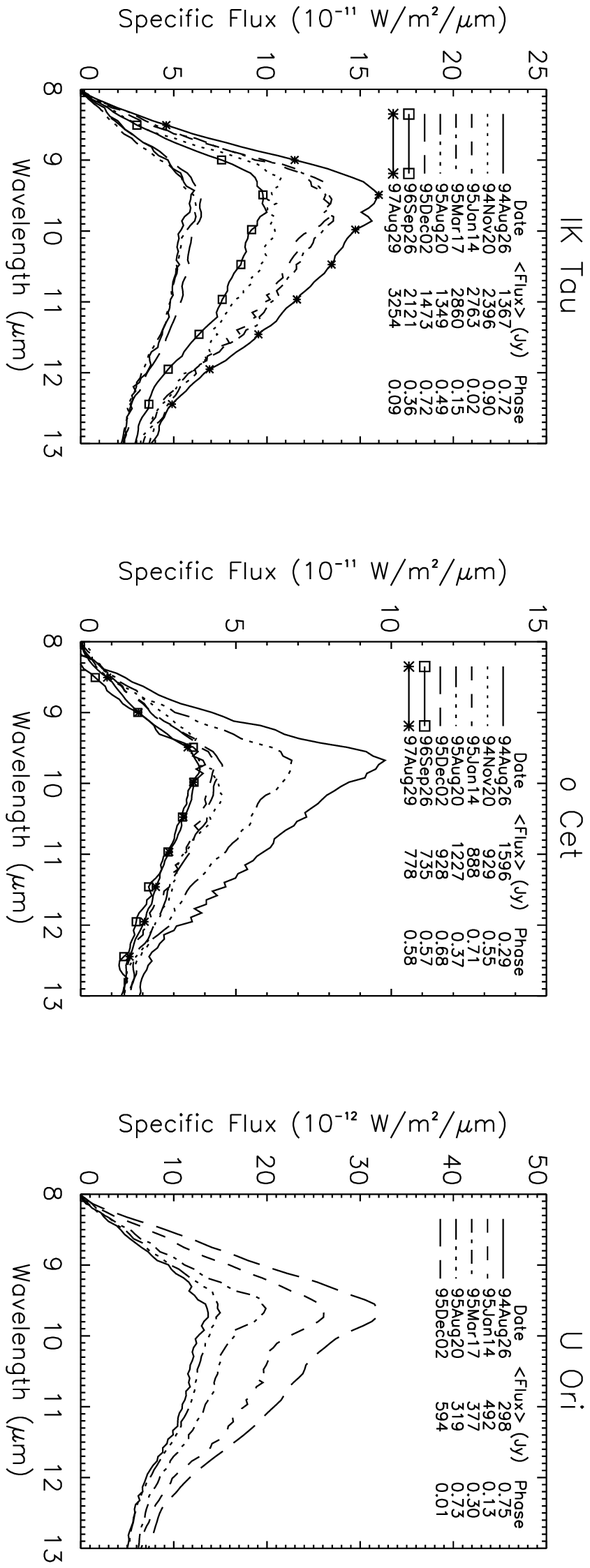}{2 in}{90.0}{75.0}{75.0}{290.0}{-150}
\caption{Stellar-subtracted, mid-infrared spectra of
IK~Tau, $o$~Cet, and U~Ori.  By subtracting an estimate of the
underlying stellar spectrum, changes in the dust spectrum itself can be
more easily seen.  Each panel includes a table containing the dates of
observation, mean (8-13\,$\micron$) fluxes of the extracted dust
spectra (in Janskys), and the pulsational phases.  Note the prominent,
narrow spectral feature near the silicate peak of U~Ori which accompanies the
star's rise towards maximum light (see \S3.2).}
\end{figure}

\begin{figure}
\figurenum{5}
%\plotone{figure5.eps}
\plotfiddle{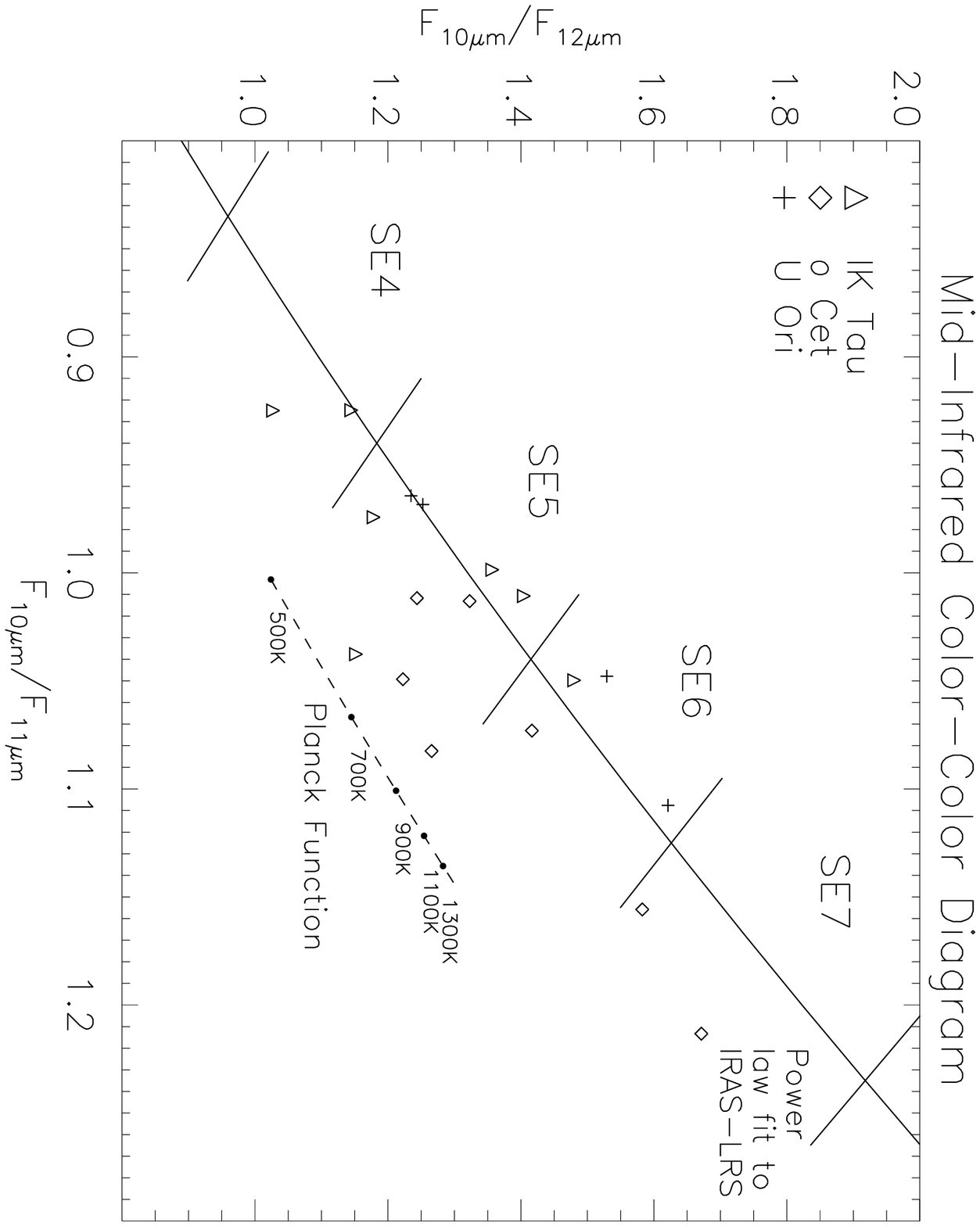}{4 in}{90.0}{75.0}{75.0}{280.0}{0}
\caption{Color-color diagram for the evolution of the
dust spectra of IK~Tau, $o$~Cet, and U~Ori.  Data points in the
lower-left portion of the diagram represent stars near minimum light.
The solid line is the power-law fit to the entire IRAS-LRS atlas
performed by Sloan \& Price (1995, SP95).  The plot is divided into
regions of similar spectral emission index (SE4-7) as defined by SP95,
which parameterizes the narrowness of the silicate feature.  The dashed
line shows the effect of dust shell temperature changes on a given
spectrum's location on this diagram.  See \S3.2 for more details.}
\end{figure}

\begin{figure}
\figurenum{6}
%\plotone{figure6.eps}
\plotfiddle{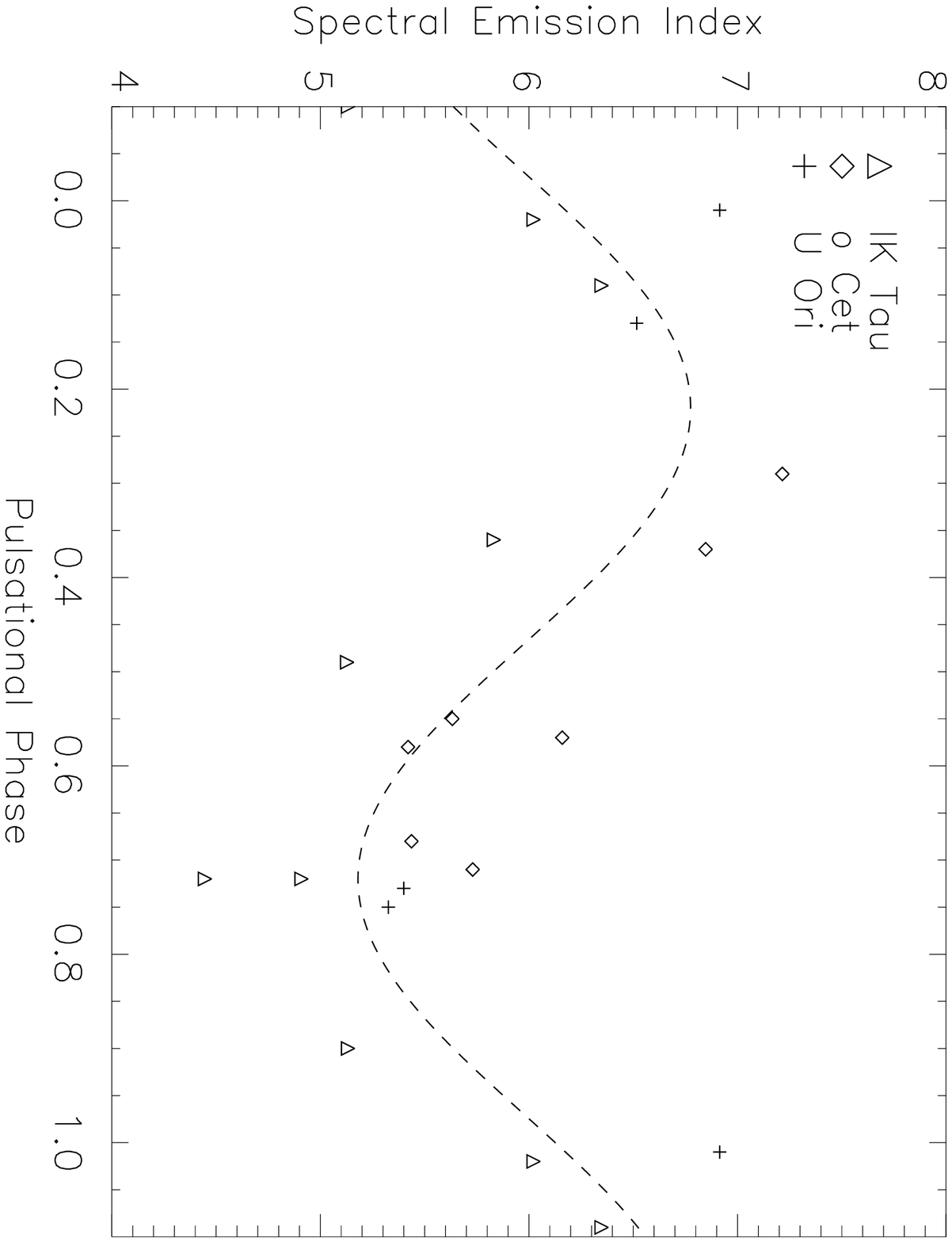}{4 in}{90.0}{75.0}{75.0}{280.0}{0}
\caption{The spectral emission (SE) index as a function of
pulsational phase for IK~Tau, $o$~Cet, and U~Ori. These data were
extracted from Figure~5 by projecting each spectrum's location on the
SP95 color-color diagram onto the power-law fit to the IRAS-LRS atlas.
The SE index, which parameterizes the narrowness of the silicate
feature, is seen to vary in phase with the pulsational cycle, reaching
maximum values (sharpest silicate features) near maximum light.  SP95
define the SE index as $10 (F_{11}/F_{12}) - 7.5$.  The dashed sinusoid
is a rough fit to the data, included only to help guide the eye.}
\end{figure}

\end{document}